\documentstyle[epsf]{aipproc}

\def\thefootnote{\fnsymbol{footnote}}
\def\bea {\begin{eqnarray}}
\def\eea {\end{eqnarray}}
\def\be {\begin{equation}}
\def\ee {\end{equation}}
\def\ben{\begin{enumerate}}
\def\een{\end{enumerate}}
\def\bi{\begin{itemize}}
\def\ei{\end{itemize}}
\def\ie{{\it i.e.}}

\def\etal{{\it et al.}}

\def\F{{\cal F}}

\def\prl {Phys. Rev. Lett.\ }
\def\pl {Phys. Lett.\ }
\def\pr {Phys. Rev.\ }
\def\np {Nucl. Phys.\ }

\def\GV{G_{\mbox{\tiny V}}}

\def\DRV{\Delta_{\mbox{\tiny R}}^{\mbox{\tiny V}}}

\def\mA{m_{\mbox{\tiny A}}}

\def\mZ{m_{\mbox{\tiny Z}}}

\def\mids{\! \mid \! }

\begin{document}

\title{Superallowed Fermi Beta Decay} 
\author{J.C. Hardy$^{a}$ and I. S. Towner$^{b}$ }
\address{$^{a}${Cyclotron Institute, Texas A \& M University, \\
College Station, TX 77843} \\
$^{b}${Physics department, Queen's University, \\
Kingston, Ontario K7L 3N6, Canada}}
\maketitle

\begin{abstract}
     Superallowed $0^+ \rightarrow 0^+$ nuclear
beta decay provides a direct measure of the weak
vector coupling constant, $\GV$.  We survey
current world data on the nine accurately
determined transitions of this type, which range
from the decay of $^{10}$C to that of $^{54}$Co,
and demonstrate that the results confirm
conservation of the weak vector current (CVC) but
differ at the 98\% confidence level from the
unitarity condition for the Cabibbo-Kobayashi-Maskawa 
(CKM) matrix.  We examine the
reliability of the small calculated corrections that
have been applied to the data, and assess the
likelihood of even higher quality nuclear data
becoming available to confirm or deny the
discrepancy.  Some of the required experiments
depend upon the availability of intense radioactive
beams.  Others are possible today.

\end{abstract}

\renewcommand{\thefootnote}{\#\arabic{footnote}}
\setcounter{footnote}{0}

\section*{Current status of world data} \label{s:ist:intro}

Superallowed Fermi $0^+ \rightarrow 0^+$ nuclear beta decays
\cite{ist:Ha90,ist:ENAM95}
provide both the best test of the Conserved Vector Current (CVC)
hypothesis in weak interactions and, together with the muon
lifetime, the most accurate value for the up-down quark-mixing
matrix element of the Cabibbo-Kobayashi-Maskawa (CKM) matrix,
$V_{ud}$. At present, the value of $V_{ud}$ deduced from nuclear
beta decay is such that, with standard values
\cite{ist:PDG98} of the other elements of the CKM matrix, the
unitarity test from the sum of the squares of the elements in the
first row fails to meet unity by more than twice the estimated
error.

According to CVC, the measured $ft$-values for Fermi
decays closely reflect the value of the weak vector coupling
constant, $\GV$, and are independent of nuclear structure,
outside of small correction terms that are of order $1\%$. 
Specifically for an
isospin-1 multiplet

\be
\F t = ft (1 + \delta_R )(1 - \delta_C) = \frac{K}{2 {\GV^{\prime}}^2},
\label{ist:Ft}
\ee

\noindent where $f$ is the statistical rate function, $t$ the partial
half-life for the transition, $\delta_R$ is the calculated nucleus-dependent
radiative correction, $\delta_C$ the calculated isospin-breaking correction,
and $K$ is a known \cite{ist:Ha90} constant.  The effective coupling
constant relates to the primitive one via
$\GV^{\prime} = \GV (1 + \DRV )^{1/2}$,
where $\DRV$ is a calculated nucleus-independent radiative
correction.  For tests of the CVC hypothesis it is not necessary to
consider this correction.

\begin{figure}[t]
\centerline{   
\epsfxsize=14cm
\epsfbox{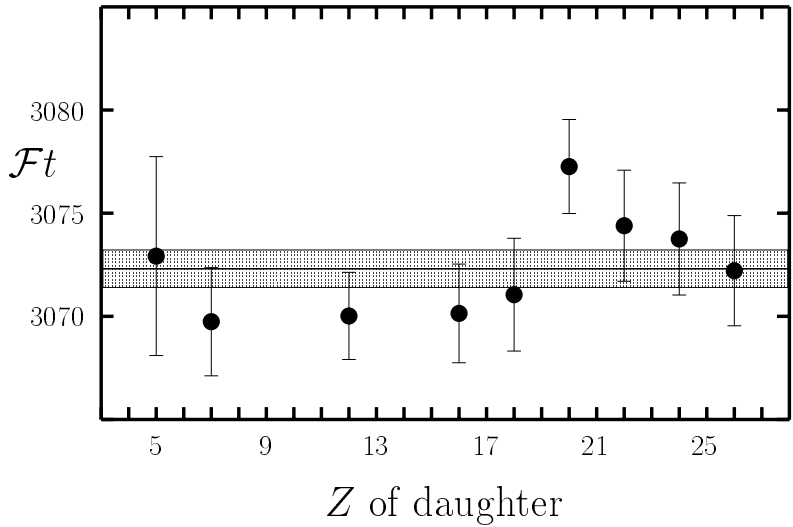}
}
\vspace{-9.5cm}
\caption{$\F t$-values for the nine precision data, and the best
least-squares one-parameter fit  \label{ist:fig1}}
\end{figure}

World data on $Q$-values, lifetimes and branching ratios were
thoroughly surveyed \cite{ist:Ha90} in 1989 and updated 
again \cite{ist:ENAM95}
for the ENAM95 conference.  Since then, there has been
a new $^{10}$C branching-ratio 
measurement \cite{ist:Fr98} and a more precise
$^{38m}$K $Q$-value determination \cite{ist:Ba98}.
We have incorporated both measurements into our data
base and extracted the $\F t$-values plotted in 
Fig.\ \ref{ist:fig1}, which also uses the $\delta_R$ and
$\delta_C$ corrections tabulated in our ENAM95 report
\cite{ist:ENAM95}.  It should be noted that those values of
$\delta_C$ are, in fact, the averages of two independent
calculations \cite{ist:THH77,ist:OB95}.  In a real sense,
both experimentally and theoretically,  Fig.\ \ref{ist:fig1}
represents the totality of current world knowledge.  The
uncertainties shown reflect the experimental uncertainties
and an estimate of the {\em relative} uncertainties in
$\delta_C$.  There is no statistically significant evidence of
inconsistencies in the data ($\chi^2/\nu = 1.1$),
thus verifying the expectation of CVC at the level of
$3 \times 10^{-4}$, the fractional uncertainty quoted on the
average $\F t$-value ($3072.3 \pm 0.9$ s).

 In using the average
$\F t$-value to determine $V_{ud}$ and test CKM unitarity
it is important to incorporate the `systematic' uncertainty in
$\delta_C$ that arises from the small systematic differences
between the two independent model calculations
\cite{ist:THH77,ist:OB95}.
 The result is

\be
\F t = 3072.3 \pm 2.0 ~{\rm s}.
\label{ist:avgFt}
\ee

\noindent With this value, an estimate \cite{ist:Si94} of the
nucleus-independent radiative correction of $\DRV = (2.40 \pm 0.08)\%$,
and the weak vector coupling constant \cite{ist:PDG98} derived from
muon decay,
we obtain

\be
V_{ud} = 0.9740 \pm 0.0005 .
\label{ist:Vud}
\ee

\noindent The quoted uncertainty is dominated by uncertainties
in the theoretical
corrections, $\DRV$ and $\delta_C$.  On adopting the 
values \cite{ist:PDG98}
of $V_{us}$
and $V_{ub}$ from the Particle Data Group, the sum
of squares of the elements in the first row of the CKM matrix,

\be
\mids V_{ud} \mids^2 +
\mids V_{us} \mids^2 +
\mids V_{ub} \mids^2 = 0.9968 \pm 0.0014,
\label{ist:unit}
\ee

\noindent differs from unity at the $98\%$ confidence level.

To restore unitarity, the calculated radiative corrections
would have to be shifted downwards by 0.3\% ($\ie$ as
much as one-quarter of their current value), or the calculated
Coulomb correction shifted upwards by 0.3\% (nearly 
one-half their value), or some combination of the two.  In what
follows, we discuss the accuracy of these two corrections
and the direction of future research.

\section*{Radiative corrections}
\label{s:ist:rc}

As mentioned, the radiative correction is conveniently divided into
terms that are nucleus-dependent, $\delta_R$, and terms that are not,
$\DRV$.  These are written

\bea
\delta_R & = & \frac{\alpha}{2 \pi} \left [ \overline{g}(E_m)
+ \delta_2 + \delta_3 + 2 C_{NS} \right ]
\nonumber \\
\DRV & = & \frac{\alpha}{2 \pi} \left [ 4 \ln (\mZ /m_p ) + \ln (m_p / \mA )
+ 2 C_{\rm Born} \right ] + \cdots ,
\label{ist:drDR}
\eea

\noindent where the ellipses represent further small terms of order
0.1\%.  In these equations, $E_m$ is the maximum electron energy
in beta decay, $\mZ$ the $Z$-boson mass, $\mA$ the $a_1$-meson mass, and
$\delta_2$ and $\delta_3$ the order $Z \alpha^2$ and $Z^2 \alpha^3$
contributions.  The electron-energy dependent function, $g(E_e,E_m)$
was derived by Sirlin \cite{ist:Si67}; it is here averaged over the
electron spectrum to give $\overline{g}(E_m)$.

Typical values are

\be
\delta_R \simeq 0.95 + 0.43 + 0.05 + (\alpha /\pi ) C_{NS} \% ,
\label{ist:tdr}
\ee

\noindent where $(\alpha /\pi )C_{NS}$ is of order $-0.3\%$ for
$T_z = -1$ beta emitters, $^{10}$C and $^{14}$O, and of order five
times smaller for the $T_z = 0$ emitters, ranging from
$-0.09\%$ to $+0.03\%$ \cite{ist:To94}.  Thus for $T_z=0$ emitters
$\delta_R \simeq 1.4\%$.  If the failure to obtain unitarity in the
 CKM matrix with
$V_{ud}$ from nuclear beta decay is due to the value of $\delta_R$,
then $\delta_R$ must be reduced to 1.1\%.  This is not likely.
The leading term, 0.95\%, involves standard QED and is well
verified.  The order-$Z\alpha^2$ term, 0.43\%, while less secure
has been calculated twice \cite{ist:Si87,ist:JR87} independently, with
results in accord.

For the nucleus-independent term

\be
\DRV = 2.12 - 0.03 + 0.20 + 0.1\% ~~\simeq~~ 2.4\%
\label{ist:tdrv}
\ee

\noindent of which the first term, the leading logarithm, is unambiguous.
Again, to achieve unitarity of the CKM matrix, $\DRV$ would have to be 
reduced to 2.1\%, \ie\ all terms other than the leading logarithm
summing to zero.  This also seems unlikely.

\section*{Coulomb corrections}
\label{s:ist:cc}

Because the leading terms in the radiative corrections are
well founded, attention has focussed more on the Coulomb
correction.  Although smaller than the radiative correction,
the Coulomb correction is clearly sensitive to
nuclear-structure issues.  It comes about because Coulomb
and charge-dependent nuclear forces destroy isospin
symmetry between the initial and final states in
superallowed beta-decay.  The consequences are twofold:
there are different degrees of configuration mixing in the
two states, and, because their binding energies are not
identical, their radial wave functions differ.  Thus we
accommodate both effects by writing $\delta_C =
\delta_{C1} + \delta_{C2}$.  Constraints can be placed on
the calculation of $\delta_{C1}$ by insisting that the
calculation reproduce the coefficients of the isobaric mass
multiplet equation.  Constraints on $\delta_{C2}$ follow
by insisting that the asymptotic forms of the proton and neutron
radial functions match known separation energies.

Recently Ormand and Brown (OB) \cite{ist:OB95} have recomputed their 
Hartree-Fock calculations with new results increasing $\delta_C$ over
their earlier work \cite{ist:OB89} but still with values systematically
smaller than the Saxon-Woods calculations of Towner, Hardy and Harvey
(THH) \cite{ist:THH77} .  Another recent work by Sagawa, van Giai and
Suzuki \cite{ist:SVS96} add RPA correlations to a Hartree-Fock
calculation; these correlations, in essence, introduce a coupling
to the isovector monopole giant resonance.  This calculation,
however, is not constrained to reproduce known separation energies.
Finally a large shell-model calculation has been mounted for the
$A=10$ case by 
Navr\'{a}til, Barrett and Ormand \cite{ist:NBO97} .  Both of these two
new works \cite{ist:SVS96,ist:NBO97} have 
produced values of $\delta_C$
{\em smaller} than those used before, \ie\ worsening rather
than helping the unitarity problem.

The typical value of $\delta_C$ is of order 0.4\%.  If the unitarity
problem is to be solved by improvements in $\delta_C$, then
$\delta_C$ has to be raised to around 0.7\%.  There is no evidence
whatsoever for such a shift from recent works.

The $\delta_C$ calculations, as pointed out by OB
\cite{ist:OB95}, do predict that $\delta_C$ should be
dramatically larger for nuclei in the $fp$-shell with $A \geq
62$.  This is due to the increasing importance of the $1p$ 
orbital, which, with its extra node in the radial function
compared to the $0f$ orbital, is much more sensitive to
Coulomb effects.  A similar effect was predicted earlier
\cite{ist:THH77} for $T_z = -1$ nuclei in the middle of the
$sd$-shell where the $1s$ orbital plays an equivalent role. 
Future experiments will test these predictions. 

\section*{Future prospects for experiment}
\label{s:ist:tfw}

The nine superallowed transitions surveyed
here have been the subject of intense scrutiny
for at least the past three decades.  All except
$^{10}$C have the special advantage that the
superallowed branch from each is by far the
dominant transition in its decay ($>$ 99\%).  This
means that the branching ratio for the 
superallowed transition can be
determined to high precision from relatively
imprecise measurements of the other weak
transitions, which can simply be
subtracted from 100\%.  Given the quantity of
careful measurements already published, are
there reasonable prospects for significant
improvements in these decays in the near
future?  Given the uncertainty in the
theoretical corrections, perhaps a more
important question is whether there is any
reason to seek experimental improvements at
all.

If we begin by accepting that it is valuable
for experiment to be at least a factor of two
more precise than theory, then an examination
of the world data shows that the Q-values for
$^{10}$C, $^{14}$O, $^{26m}$Al and
$^{42}$Sc, the half-lives of $^{10}$C,
$^{34}$Cl and $^{38m}$K, and the
branching ratio for $^{10}$C can all bear
improvement.  Such improvements will soon
be feasible.  The Q-values will reach 
the required level (and more) as mass
measurements with new on-line Penning traps
become possible; half-lives will likely yield to
measurements with higher statistics as 
high-intensity beams of separated isotopes are
developed for the new radioactive-beam
facilities; and, finally, an improved 
branching-ratio measurement on $^{10}$C has already
been made with Gammasphere and simply
awaits analysis \cite{ist:Fr98}.

Qualitative improvements will also come
as we increase the number of superallowed emitters 
accessible to precision studies.  The greatest
attention recently has been paid to the $T_z =
0$ emitters with $A \geq 62$, since these
nuclei are expected to be produced at new
radioactive-beam facilities, and their
calculated Coulomb corrections, $\delta_C$,
are predicted to be large
\cite{ist:OB95,ist:SVS96,ist:JH92}.  They could then
provide a valuable test of the accuracy of
$\delta_C$ calculations.  It is likely, though,
that the required precision will not be
attainable for some time to come.  The decays
of these nuclei will be of higher energy and
each will therefore involve several allowed
transitions of significant intensity in addition
to the superallowed transition.  Branching-ratio 
measurements will thus be very
demanding, particularly with the limited
intensities likely to be available initially for
these rather exotic nuclei.  Lifetime
measurements will be similarly constrained by
statistics.

More accessible in the short term will be
the $T_z = -1$ superallowed emitters with $18
\leq A \leq 38$.  There is good reason to
explore them.  For example, the calculated
value \cite{ist:THH77} of $\delta_C$ for $^{30}$S decay,
though smaller than the $\delta_C$'s expected
for the heavier nuclei, is actually 1.2\% -- 
about a factor of two larger than for any other
case currently known -- while $^{22}$Mg
has a very low value of 0.35\%.  If such large
differences are confirmed by the measured
$ft$-values, then it will do much to increase
our confidence in the calculated Coulomb
corrections.   To be sure, these decays will
provide a challenge, particularly in the
measurement of their branching ratios, but the
required precision should be achievable with
isotope-separated beams that are currently available. 
In fact, such experiments are already in their
early stages at the Texas A\&M cyclotron. 

\section*{Conclusions}
\label{s:ist:conc}

The current world data on superallowed $0^+
\rightarrow 0^+$ beta decays lead to a self-consistent
set of $\F t$-values that agree with CVC but differ
provocatively, though not yet definitively, from the
expectation of CKM unitarity.  There are no evident
defects in the calculated radiative and Coulomb
corrections that could remove the problem, so, if any
progress is to be made in firmly establishing (or
eliminating) the discrepancy with unitarity, additional
experiments are required.  We have indicated what
some relevant nuclear experiments might be.

Clearly, there is strong motivation to pursue them
since, if firmly established, a discrepancy with unitarity would
indicate the need for an extension of the three-generation 
Standard Model.

\end{document}